\documentclass[dvipsnames, twocolumn]{aastex63}
\usepackage[version=4]{mhchem}
\usepackage{amsmath}

\received{XX}
\revised{YY}
\accepted{ZZ}
\submitjournal{ApJ}

\shorttitle{Sub-stellar O/H and C/H and super-stellar C/O in planet feeding gas}
\shortauthors{A. D. Bosman for the MAPS collaboration}

\begin{document}
\title{Molecules with ALMA at Planet-forming Scales (MAPS). VII.

Sub-stellar O/H and C/H and super-stellar C/O in planet feeding gas}
\date{\today}

\correspondingauthor{Arthur Bosman}
\email{arbos@umich.edu}

\author[0000-0003-4001-3589]{Arthur D. Bosman}
\affiliation{Department of Astronomy, University of Michigan,
323 West Hall, 1085 S. University Avenue,
Ann Arbor, MI 48109, USA}

\author[0000-0002-2692-7862]{Felipe Alarc\'on}
\affiliation{Department of Astronomy, University of Michigan,
323 West Hall, 1085 S. University Avenue,
Ann Arbor, MI 48109, USA}

\author[0000-0003-4179-6394]{Edwin A. Bergin}
\affiliation{Department of Astronomy, University of Michigan,
323 West Hall, 1085 S. University Avenue,
Ann Arbor, MI 48109, USA}

\author[0000-0002-0661-7517]{Ke Zhang}
\altaffiliation{NASA Hubble Fellow}
\affiliation{Department of Astronomy, University of Wisconsin-Madison,
475 N Charter St, Madison, WI 53706}
\affiliation{Department of Astronomy, University of Michigan,
323 West Hall, 1085 S. University Avenue,
Ann Arbor, MI 48109, USA}

\author[0000-0002-2555-9869]{Merel L.R. van 't Hoff}
\affiliation{Department of Astronomy, University of Michigan,
323 West Hall, 1085 S. University Avenue,
Ann Arbor, MI 48109, USA}



\author[0000-0001-8798-1347]{Karin I. \"Oberg}
\affiliation{Center for Astrophysics \textbar{} Harvard \& Smithsonian, 60 Garden St., Cambridge, MA 02138, USA}

\author[0000-0003-4784-3040]{Viviana V. Guzm\'an}
\affil{Instituto de Astrof\'isica, Pontificia Universidad Cat\'olica de Chile, Av. Vicu\~na Mackenna 4860, 7820436 Macul, Santiago, Chile}

\author[0000-0001-6078-786X]{Catherine Walsh}
\affiliation{School of Physics and Astronomy, University of Leeds, Leeds LS2 9JT, UK}

\author[0000-0003-3283-6884]{Yuri Aikawa}
\affiliation{Department of Astronomy, Graduate School of Science, The University of Tokyo, Tokyo 113-0033, Japan}

\author[0000-0003-2253-2270]{Sean M. Andrews}
\affiliation{Center for Astrophysics \textbar{} Harvard \& Smithsonian, 60 Garden St., Cambridge, MA 02138, USA}

\author[0000-0002-8716-0482]{Jennifer B. Bergner} 
\altaffiliation{NASA Hubble Fellowship Program Sagan Fellow}
\affiliation{University of Chicago Department of the Geophysical Sciences, Chicago, IL 60637, USA}

\author[0000-0003-2014-2121]{Alice S. Booth}
\affiliation{Leiden Observatory, Leiden University, 2300 RA Leiden, the Netherlands}
\affiliation{School of Physics and Astronomy, University of Leeds, Leeds, LS2 9JT UK}

\author[0000-0002-2700-9676]{Gianni Cataldi}
\affil{National Astronomical Observatory of Japan, Osawa 2-21-1, Mitaka, Tokyo 181-8588, Japan}
\affil{Department of Astronomy, Graduate School of Science, The University of Tokyo, Tokyo 113-0033, Japan}

\author[0000-0003-2076-8001]{L. Ilsedore Cleeves}
\affiliation{University of Virginia, 530 McCormick Rd, Charlottesville, VA 22904, USA}

\author[0000-0002-1483-8811]{Ian Czekala}
\altaffiliation{NASA Hubble Fellowship Program Sagan Fellow}
\affiliation{Department of Astronomy and Astrophysics, 525 Davey Laboratory, The Pennsylvania State University, University Park, PA 16802, USA}
\affiliation{Center for Exoplanets and Habitable Worlds, 525 Davey Laboratory, The Pennsylvania State University, University Park, PA 16802, USA}
\affiliation{Center for Astrostatistics, 525 Davey Laboratory, The Pennsylvania State University, University Park, PA 16802, USA}
\affiliation{Institute for Computational \& Data Sciences, The Pennsylvania State University, University Park, PA 16802, USA}
\affiliation{Department of Astronomy, 501 Campbell Hall, University of California, Berkeley, CA 94720-3411, USA}

\author[0000-0002-2026-8157]{Kenji Furuya} \affiliation{National Astronomical Observatory of Japan, Osawa 2-21-1, Mitaka, Tokyo 181-8588, Japan}

\author[0000-0001-6947-6072]{Jane Huang}
\altaffiliation{NASA Hubble Fellowship Program Sagan Fellow}
\affiliation{Department of Astronomy, University of Michigan, 323 West Hall, 1085 S. University Avenue, Ann Arbor, MI 48109, USA}
\affiliation{Center for Astrophysics \textbar{} Harvard \& Smithsonian, 60 Garden St., Cambridge, MA 02138, USA}

\author[0000-0003-1008-1142]{John D. Ilee}
\affiliation{School of Physics and Astronomy, University of Leeds, Leeds LS2 9JT, UK}

\author[0000-0003-1413-1776]{Charles J. Law}
\affiliation{Center for Astrophysics \textbar{} Harvard \& Smithsonian, 60 Garden St., Cambridge, MA 02138, USA}

\author[0000-0003-1837-3772]{Romane Le Gal}
\affiliation{Center for Astrophysics \textbar{} Harvard \& Smithsonian, 60 Garden St., Cambridge, MA 02138, USA}
\affiliation{IRAP, Universit\'{e} de Toulouse, CNRS, CNES, UT3, 31400 Toulouse, France}
\affiliation{Univ. Grenoble Alpes, CNRS, IPAG, F-38000 Grenoble, France} 
\affiliation{IRAM, 300 rue de la piscine, F-38406 Saint-Martin d'H\`{e}res, France}

\author[0000-0002-7616-666X]{Yao Liu}
\affiliation{Purple Mountain Observatory \& Key Laboratory for Radio Astronomy, Chinese Academy of Sciences, Nanjing 210023, China}

\author[0000-0002-7607-719X]{Feng Long}
\affiliation{Center for Astrophysics \textbar{} Harvard \& Smithsonian, 60 Garden St., Cambridge, MA 02138, USA}

\author[0000-0002-8932-1219]{Ryan A. Loomis}
\affiliation{National Radio Astronomy Observatory, Charlottesville, VA 22903, USA}

\author[0000-0002-1637-7393]{Fran\c cois M\'enard}
\affiliation{Univ. Grenoble Alpes, CNRS, IPAG, F-38000 Grenoble, France}

\author[0000-0002-7058-7682]{Hideko Nomura}
\affiliation{National Astronomical Observatory of Japan, Osawa 2-21-1, Mitaka, Tokyo 181-8588, Japan}

\author[0000-0001-8642-1786]{Chunhua Qi}
\affiliation{Center for Astrophysics \textbar{} Harvard \& Smithsonian, 60 Garden St., Cambridge, MA 02138, USA}

\author[0000-0002-6429-9457]{Kamber R. Schwarz}
\altaffiliation{NASA Hubble Fellowship Program Sagan Fellow}
\affiliation{Lunar and Planetary Laboratory, University of Arizona, 1629 East University Boulevard, Tucson, AZ 85721, USA}

\author[0000-0003-1534-5186]{Richard Teague}
\affiliation{Center for Astrophysics \textbar{} Harvard \& Smithsonian, 60 Garden St., Cambridge, MA 02138, USA}

\author[0000-0002-6034-2892]{Takashi Tsukagoshi}
\affiliation{National Astronomical Observatory of Japan, Osawa 2-21-1, Mitaka, Tokyo 181-8588, Japan}

\author[0000-0003-4099-6941]{Yoshihide Yamato}
\affiliation{Department of Astronomy, Graduate School of Science, The University of Tokyo, Tokyo 113-0033, Japan}

\author[0000-0003-1526-7587]{David J. Wilner}
\affiliation{Center for Astrophysics \textbar{} Harvard \& Smithsonian, 60 Garden St., Cambridge, MA 02138, USA}


\begin{abstract}
The elemental composition of the gas and dust in a protoplanetary disk influences the compositions of the planets that form in it. We use the Molecules with ALMA at Planet-forming Scales (MAPS) data to constrain the elemental composition of the gas at the locations of potentially forming planets.
The elemental abundances are inferred by comparing source-specific gas-grain thermochemical models, with variable C/O ratios and small-grain abundances, from the DALI code with CO and \ce{C2H} column densities derived from the high-resolution observations of the disks of AS 209, HD 163296, and MWC 480. Elevated C/O ratios ($\sim 2.0$), even within the CO ice line, are necessary to match the inferred \ce{C2H} column densities, over most of the pebble disk. Combined with constraints on the CO abundances in these systems, this implies that both the O/H and C/H ratios in the gas are substellar by a factor of 4--10, with the O/H depleted by a factor of 20--50, resulting in the high C/O ratios. This necessitates that even within the CO ice line, most of the volatile carbon and oxygen is still trapped on grains in the midplane. Planets accreting gas in the gaps of the AS 209, HD 163296, and MWC 480 disks will thus acquire very little carbon and oxygen after reaching the pebble isolation mass. In the absence of atmosphere-enriching events, these planets would thus have a strongly substellar O/H and C/H and superstellar C/O atmospheric composition.
This paper is part of the MAPS special issue of the Astrophysical Journal Supplement.
\end{abstract}


\section{Introduction}
More than 10\% of stars are thought to host a Jupiter-sized planet \citep{Johnson2010, Winn2015}. Current observational campaigns are probing the atmospheric composition of these gas giants, with future instruments promising better characterization of more gas giants \citep[e.g.][]{Madhusudhan2014, Sing2016, Greene2016}. These composition measurements place constraints on the process of planet formation and evolution by comparison of the composition of protoplanetary disks and the planets that formed within them \citep[e.g.][]{Madhusudhan2017}. Achieving that goal will require measurements of the elemental composition of disks, particularly around substructures (e.g., gaps) in the millimeter disk, that are likely opened by young planets \citep[e.g.][]{Paardekooper2006dust, Rosotti2016, Zhang2018}.

The composition of a planet is predicted to depend on the location of formation \citep[e.g.][]{Oberg2011, Cridland2017, Oberg2020}, particularly relative to the ice/gas transitions (ice lines) of major volatile carriers such as \ce{CO}, \ce{CO2}, \ce{H2O}, \ce{N2}, and \ce{NH3}. Further studies have shown that the transport of ices via sedimentation and inward drift of pebbles also has a significant impact on the composition of the gas and ice that can be accreted by planets \citep{Ciesla2006, Piso2015, Booth2017, Booth2019, Krijt2020}. All of these studies assume that the disk material has a volatile interstellar-medium (ISM) elemental composition with various additional assumptions regarding chemistry and disk dynamics. The model-to-model variations create significant uncertainties for predicting planetary composition \citep[e.g.][]{Eistrup2016, Booth2017}.

A more direct link between the composition and formation location of a giant planet could be made by comparing the elemental composition of the material at the specific radial locations or sites of planet formation with present-day gas giant exoplanet atmospheric composition. On top of this, a measurement of the elemental composition of the disk during planet formation can act as an intermediary calibration point to improve the accuracy of population synthesis models \citep[e.g.,][]{Mordasini2009}.

Accurately measuring the elemental composition of the disk requires an observational data set that covers many different molecular species, as well as a physical understanding of the regions being probed by the observations \citep[e.g.][]{Bergin2016, Trapman2017, Cleeves2018, Bosman2019}. This limits the derivation of elemental composition to disks with a broad range of observations and with a good physical model available. With the Molecules with ALMA at Planet-forming Scales (MAPS) data and ancillary modeling efforts \citep{oberg20, zhang20}\footnote{\url{http://www.alma-maps.info}}, the disks in the MAPS sample are perfect objects for an elemental composition study, especially as the high-resolution of the data allows for a radially resolved measurement of the elemental composition.

In this paper, we will focus on the carbon and oxygen abundances, which are derived by comparison of models directly to the high-resolution MAPS data that resolve putative sites of incipient giant planet formation. The carbon and oxygen abundances in the disk will be derived using constraints on the CO column from \citet{zhang20} and the \ce{C2H} column from \citet{guzman20}. CO is used to constrain the available total C or O column (e.g., C/H and O/H), and we use the known dependency of \ce{C2H} on the carbon-to-oxygen (C/O) ratio to determine the C/O ratio in the disk gas \citep[e.g.][]{Bergin2016, Miotello2019, Bosman2021}.
A focus on the C/O ratio in the disk gas is well matched to the available volatile abundance measurements of exoplanetary atmospheres \citep{Madhusudhan2014, Greene2016, Madhusudhan2019}.\footnote{We note that nitrogen abundance can also be indicative of a planet's origin \citep{Cridland2017}, and that it is possible to constrain the nitrogen abundance within protoplanetary disks \citep{Cleeves2018}, which will be the subject of future work.}

\section{Methods}

\subsection{Source selection}

\begin{figure*}
    \centering
    \includegraphics[width = \hsize]{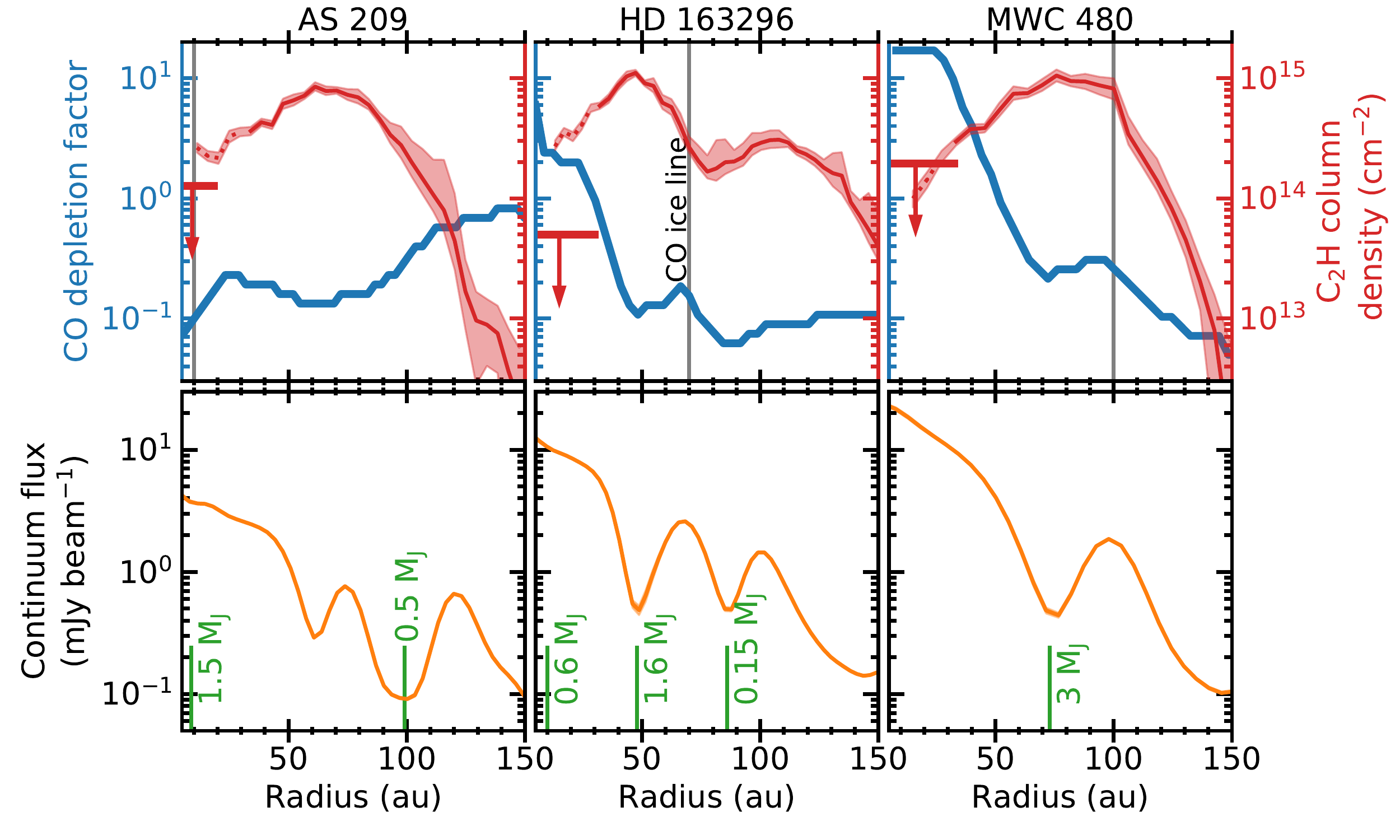}
    \caption{\textit{Top: }Comparison of the CO depletion (blue, left axis) from \citet{zhang20}, with the \ce{C2H} column (red, right axis) derived in \citet{guzman20}, for, (from left to right) the AS 209, MWC 480 and HD 163296 disks. The solid line shows where the \ce{C2H} column density derivation is assumed to be robust. In the inner 30 au the imagecube based fitting technique is less robust and the column density derived values for these radii are shown with a dotted line. For each disk the upper limit to the \ce{C2H} abundance in the inner disk as derived from the line profiles is also shown (See Appendix~\ref{app:C2H_col_deriv})).
    The grey vertical line shows the location of the CO ice line.
    \textit{Bottom:} Continuum azimuthal average radial intensity profile from \citet{sierra20} at 220 GHz. 
    The green vertical lines show the locations of planets inferred from the millimeter dust structure, together with their inferred mass \citep[all planet masses assume a viscous alpha of 10$^{-3}$,][]{Zhang2018, Liu2019}. }
    
    \label{fig:CO_C2H_comp}
\end{figure*}

We will focus on three of the five disks targeted by MAPS: the disks around AS 209, HD 163296, and MWC 480. All three of these disks have deep gaps in the millimeter emission indicative of ongoing planet formation \citep[see Fig.~\ref{fig:CO_C2H_comp};][]{Andrews2018, Huang2018, Zhang2018, Guzman2018, Long2018, Liu2019}, as well as gas velocity structures thought to be perturbations induced by unseen planets in these gaps \citep{Pinte2018, Teague2018, Teague2019, Teague20}. We leave out the disks around IM Lup and GM Aur. The composition of the IM Lup disk has already been discussed in detail in \citet{Cleeves2018}. The GM Aur disk is accreting new material, influencing the chemical composition of the disk \citep{huang20, schwarz20}, complicating modeling of the disk.

Furthermore, in these disks, we will focus on the gas on the scale of the millimeter disk, that is, the size of the disk in the millimeter continuum, using these large grains as a signpost of active planet formation. Fig.~\ref{fig:CO_C2H_comp} shows millimeter continuum radial profiles, rich in structure, for the three disks that are studied together with the planet masses that are inferred to cause these structures \citep{Zhang2018, Liu2019, sierra20}\footnote{All planet masses assume a viscous $\alpha$ \citep[][]{Shakura1973} of 10$^{-3}$.} All three disks show potentially planet-induced dust structure that is clearly seen at with the resolution of the MAPS continuum data \citep[$\sim$0$\farcs$1 $\times$ 0$\farcs$1][]{sierra20}. The MAPS data are thus able to distinguish the gap region from the surrounding disk material.

{\em Our focus is understanding the chemical composition of localized regions associated with incipient planet formation.}
In the AS 209 disk, the structures of interest are the gaps at 61 and 105 au, straddling a bright ring at 74 au \citep{Guzman2018}.  These are all well outside of the CO ice line, which is located around 15 au \citep{zhang20, alarcon20}. In the HD 163296 disk, we will focus on the dust gaps at 48 and 86 au, with one gap inside and one gap outside the ice line at 70 au \citep{Isella2018, Zhang2020a, zhang20, Calahan20b}. Finally, for the MWC 480 disk we focus on the 74 au gap, which is inside the CO ice line at 100 au \citep{Liu2019, zhang20}. 
We disregard the inner $\sim$10 au, as an accurate \ce{CO} column density has not been derived for this region \citep{zhang20}.

To constrain the oxygen and carbon elemental abundances in the three selected disks we base ourselves on results and models developed within the MAPS ALMA large program \citep{oberg20}. For \ce{C2H} the column density derived in \citet{guzman20} is used. These column densities are derived from the imagecube by fitting the different \ce{C2H} hyperfine components.
The \ce{C2H} column density derived from the images is uncertain within 30 au. As such we derive a separate constraint on the \ce{C2H} column within 30 au using the kinematic information in the line profile. This is discussed in Appendix~\ref{app:C2H_col_deriv}. The shown upper limits in Fig.~\ref{fig:CO_C2H_comp} assume a column of \ce{C2H} that is constant in the inner 20 (AS 209) or 30 (HD 163296 and MWC 480) au.

We adopt the CO depletion factor \citet{zhang20} derived from the observed CO isotopologues radial profiles \citep{law20_rad}. \citet{zhang20} build disk-specific models based on high-resolution continuum and CO isotopologue ($^{13}$CO and C$^{18}$O) line data. A total gas-to-dust mass ratio of 100 is assumed, with the gas disk size set by the extent of CO isotopologue emission. Gas and small dust (0.005 $\mu$m - 1$\mu$m) are well coupled and have the standard tapered-power law surface density distribution and a vertical Gaussian density distribution \citep[see][for the disk parameters]{zhang20}. The large dust is settled, at 0.2 times the gas scale height and is structured to match the observed millimeter structure \citep[see, ][Tables 1 and 2 and Fig. 5]{zhang20}. The CO abundance in the disk is then varied until the model CO isotopologue line emission matches the observed line emission for  \ce{^{13}CO} and \ce{C^{18}O} for both the $J=$2-1 and $J=$1-0 transitions. The ratio between the abundance necessary to reproduce that data and the ISM CO abundance is called the CO depletion factor. The model CO columns are consistent with the CO column derived from the hyperfine lines of \ce{C^{17}O} in the regions the latter is detected. Both the \ce{C2H} column and the CO depletion factor are shown in Fig.~\ref{fig:CO_C2H_comp}.

\subsection{\ce{C2H} chemistry}
The chemistry of \ce{C2H} and its use as a tracer of the elemental C/O ratio in protoplanetary disks has been discussed for a variety of sources and conditions \citep[e.g.][]{Jansen1995, Nagy2015, Bergin2016, Cleeves2018, Bosman2021}.
As a radical it is most abundant in regions with an active photo-chemistry, where it is continuously produced and destroyed.
The chemistry of \ce{C2H} is tightly linked to that of \ce{CH4} and \ce{C2H2}, which is the main reservoir of carbon that is not locked in CO.
\ce{C2H} can be directly formed by photo-dissociation of \ce{C2H2} or it is formed by dissociative recombination of the \ce{C2H2+} and \ce{C2H3+} ions, which are part of a gas-phase hydrocarbon radical and ion cycle \citep[see, for example, Fig.~1 of ][]{Bosman2021}.

The \ce{C2H} abundance is very dependent on the C/O ratio in the gas, with reactions as \ce{C2H + O -> CO + CH} and \ce{CH + O -> CO + H}, directly driving down the \ce{C2H} abundance as well as removing carbon from the hydrocarbon reservoir. Reactions generally result in the formation of CO, which is stable against photo-dissociation \citep{Dishoeck1988}. As such this carbon is no longer available for hydrocarbon chemistry. Depending on the physical structure of the disk and the stellar radiation field, the \ce{C2H} column density in a disk model can vary by four orders of magnitude, just by variations of the C/O ratio between $\sim$0.5 and 2.0 \citep{Bergin2016, Cleeves2018}. This makes \ce{C2H} a very good tracer of the C/O ratio.

As a radical \ce{C2H} traces the photo-dissociation region (PDR) layers of the disk. At low C/O ratios, all carbon for \ce{C2H} comes from the dissociation of \ce{CO}, which happens only in the top layer of the disk PDR, where CO self-shielding is not yet active. The mass in this thin layer is relatively constant in a disk, as well as disk-to-disk as it depends on a fundamental property of the CO molecule, which starts to self-shield at a column of $10^{15}$ cm$^{-2}$ \citep{Dishoeck1988}. At C/O ratios above 1  the carbon for the formation of \ce{C2H} can be extracted from hydrocarbons as \ce{CH4} and \ce{C2H2}. The thickness of the layer in which these hydrocarbons can be dissociated depends on UV attenuation by dust. The small dust content of the disk atmosphere thus has an effect on the measured \ce{C2H} column. Initial calculations in \citet{Bosman2021} predict that the small dust content will impact the \ce{C2H} abundance, but is relatively weak. An order of magnitude drop in small dust abundance should only lead to an increase in the \ce{C2H} abundance by a factor two.  

\subsection{Modeling setup}
In our modeling we start with the model structure and CO depletion factor as derived by \citet{zhang20}. We vary the C/O ratio and the amount of small dust in the surface layers as these parameters have the most impact of the \ce{C2H} abundance.
The small dust content of the disks is varied by multiplying the small dust abundances by a constant factor of 0.001, 0.01, 0.1 or 1 from the base models of \citet{zhang20}.   The last of these are the fiducial dust structures and are identical to those used in \citet{zhang20}.\footnote{For models with lower small-grain abundances the model} SED no longer fits the observational SED. Each model has a constant C/O ratio of either 0.47, 1.0, or 2.0. These correspond to the volatile ISM level (0.47), CO dominated gas (1.0) and a high C/O value (2.0). Increasing the C/O ratio above 2.0 no longer has an impact of the \ce{C2H} abundance \citep[e.g.,][]{Cleeves2018}. Combining the four different small dust abundances and 3 different C/O ratios this results in twelve models per disk studied.

The model calculation is done in two steps. In the first step the dust and gas temperatures are calculated. We start from the best fit structure and CO depletion factors from \citet{zhang20}. The structures are here modified to vary the small dust content. The abundances of \ce{H2O} and \ce{CH4} in initial conditions of the chemical network are also modified at this point so the gas temperature calculation includes the changes in composition driven by the different C/O ratios. The dust and gas temperatures are then calculated using DALI with the standard reduced chemical network, which correctly calculates the abundances of the major gas coolants. \citep{Bruderer2012,Bruderer2013}. In the second step, a full gas-grain network was used to calculate the 2D-chemical composition based on the physical conditions determined in the first step.

This two step approach was chosen to balance consistency with feasibility. To calculate the gas temperature, the chemical solver has to run multiple times per cell, which is prohibitively expensive to do with the gas-grain chemical network. As mentioned, this approach does capture the effects of the C/O ratio on the gas temperature, and the effect of these changes on the resulting chemistry, while being computationally tractable.

\subsection{Gas-grain chemical network}

The gas-grain chemical network used for the second step of our calculations is based on the network of \citet{Walsh2015}, which is a combination of the \textrm{UMIST12}\footnote{\url{http://udfa.ajmarkwick.net/}} \citep{McElroy2013} for the gas-phase reactions and the Ohio State University (OSU) network \citep{Garrod2008gas-grain} for the grain surface reactions. Molecular binding energies are from \citet{McElroy2013} and \citet{Penteado2017}, with small adjustments for \ce{NH_x} and \ce{CH_x} radicals \citep{Bosman2018CO}. These binding energies are tabulated in \citet{Bosman2018CO}, Table B.2. The freeze-out of \ce{H2} is also treated as in \citet{Bosman2018CO}, using the higher (430 K) binding energy for \ce{H2} when there are less than two layers of \ce{H2} ice, but lowering the binding energy to 100 K when \ce{H2} covers the ice completely.

Photo-desorption yields are taken from experiments where available \citep{Oberg2009CH3OH, Oberg2009H2O, Oberg2009CO_N2_CO2}, and assumed to be $10^{-3}$ otherwise. All photodesorption is assumed to yield intact molecules in the gas-phase.\footnote{Many molecules are expected to fragment on absorption of a UV photon, with some of the fragments released from the ice, instead of photo-desorbing intact as assumed. Fragmentation has been shown for methanol \citep{Bertin2016} and is expected for other, larger molecules.} This is not expected to impact the chemistry of the species studied here. The grain surface chemistry assumes a diffusion-to-binding energy ratio of 0.3 and a tunneling barrier of 1\,\AA{} \citep[as in][]{Bosman2018CO}. For photo-dissociation, the wavelength dependent cross-sections of \citet{Heays2017} are used in combination with the wavelength dependent UV field that DALI calculates. For the few cases in which \citet{Heays2017} do not have a compilation the wavelength dependent cross sections, the rate parameterization of \citet{McElroy2013} is used. Self-shielding and mutual shielding is included for \ce{C}, \ce{CO}, \ce{N2} and \ce{H2} \citep{Draine1996, Kamp2000, Visser2009, Visser2018}.

The initial abundances used in the models with different  C/O ratios are tabulated in Table~\ref{tab:init_abu}. Any molecule, atom, or ion not listed in the table has zero abundance at the beginning of the chemical calculation. The abundances of \ce{H2O}, \ce{CO}, and \ce{CH4}, that is, the initial carbon and oxygen carriers, are scaled with the CO depletion factors derived in \citet{zhang20} (see Fig.~\ref{fig:CO_C2H_comp}). Further the elemental abundances have been choses such that the maximum amount of CO that can be form, that is the lowest of the carbon and oxygen elemental abundances, is the same for all three of the C/O ratio. \ce{CH4} was chosen as the carrier for the excess carbon (instead of elemental C) in the models. The exact initial abundances do not matter greatly for this calculation, as abundances in the \ce{C2H} layer will be set by photon-dominated steady-state chemistry \citep{Bosman2021}. The only exceptions to this rule are the large (more than seven carbon atoms) carbon chains, which act as refractories in the intermediate PDR layers. These large carbon chains are only formed in the models when a significant fraction ($>$10\%) of the carbon is in elemental or ionized form. To prevent this, elemental or ionized form should not be chosen for the initial carbon carrier. In other cases, the exact carrier of the excess carbon does not substantially alter the resulting \ce{C2H} abundance \citep{Bosman2021}.

A low cosmic-ray ionization rate of $10^{-18}$ s$^{-1}$ is used for the chemical models, similar to the models in \citet{zhang20} \citep[see, e.g.][for a discussion of cosmic ray propagation in disks.]{Cleeves2013}. A higher cosmic-ray ionization rate, coupled with the already low CO abundances currently assumed, would quickly convert the remaining CO into other species like \ce{CH4}, \ce{CO2} and \ce{CH3OH} \citep[e.g.][]{Furuya2014, Schwarz2016, Bosman2018CO} in the 1 Myr chemical timescale that is assumed. This would create a disagreement between the CO column in the models and the intended, observationally motivated CO column density. This cosmic-ray ionization rate is thus not the rate necessary to lower the CO abundance from the ISM level to the observed levels, but it is the cosmic-ray ionization rate that has to be used to prevent the CO abundance from dropping further. We note that the cosmic-ray ionization rate used does not impact the \ce{C2H} abundance in chemical models \citep{Cleeves2018} and that the cosmic-ray ionization rate used here is consistent with modeling of MAPS sources by \citet{aikawa20}.

Appendix~\ref{app:CO_cols} shows the CO columns for difference C/O ratios for the three disks compared to the CO columns derived from the data. The difference C/O ratios do not impact the CO column in the model. Furthermore, the agreement between our model CO columns and the results from \citep{zhang20} indicates that the amount of volatiles in the surface layer is representative for the disks studied.

\begin{table}[]

    \caption{Initial abundances for the chemical models}
    \label{tab:init_abu}
    \centering
    \begin{tabular}{l r r r}
    \hline
    \hline
        Species &  C/O = 0.47 & C/O = 1.0 & C/O = 2.0\\
        \hline
        \ce{H2} & 0.495& 0.495& 0.495\\
        \ce{H} & 0.01& 0.01& 0.01\\
        \ce{H2O}$^*$ & $1.53\times 10^{-4}$& $10^{-10}$& $10^{-10}$\\
        \ce{CO}$^*$ & $1.35\times 10^{-4}$& $1.35\times 10^{-4}$& $1.35\times 10^{-4}$\\
        \ce{CH4}$^*$ & $10^{-10}$& $10^{-10}$&$1.35\times 10^{-4}$\\
        \ce{N2} & $1.07\times 10^{-5}$ & $1.07\times 10^{-5}$ &$1.07\times 10^{-5}$ \\
        \ce{S} & $1.91\times 10^{-8}$& $1.91\times 10^{-8}$& $1.91\times 10^{-8}$\\
        \ce{Si} & $7.94\times 10^{-8}$ & $7.94\times 10^{-8}$ &$7.94\times 10^{-8}$\\
        \ce{Fe} & $4.27\times 10^{-9}$& $4.27\times 10^{-9}$& $4.27\times 10^{-9}$\\
        \ce{Mg} & $1.00\times 10^{-11}$& $1.00\times 10^{-11}$& $1.00\times 10^{-11}$\\
        \hline
    \end{tabular}
    \tablecomments{$^*$ The \ce{H2O}, \ce{CO} and \ce{CH4} abundances shown here assume a CO depletion factor of 1.0, The actual abundance of these species at each radius is obtained by multiplying these abundances the CO depletion factor of Fig.~\ref{fig:CO_C2H_comp}}

\end{table}

\section{Composition of the disk gas}

\begin{figure*}
    \centering
    \includegraphics[width=\hsize]{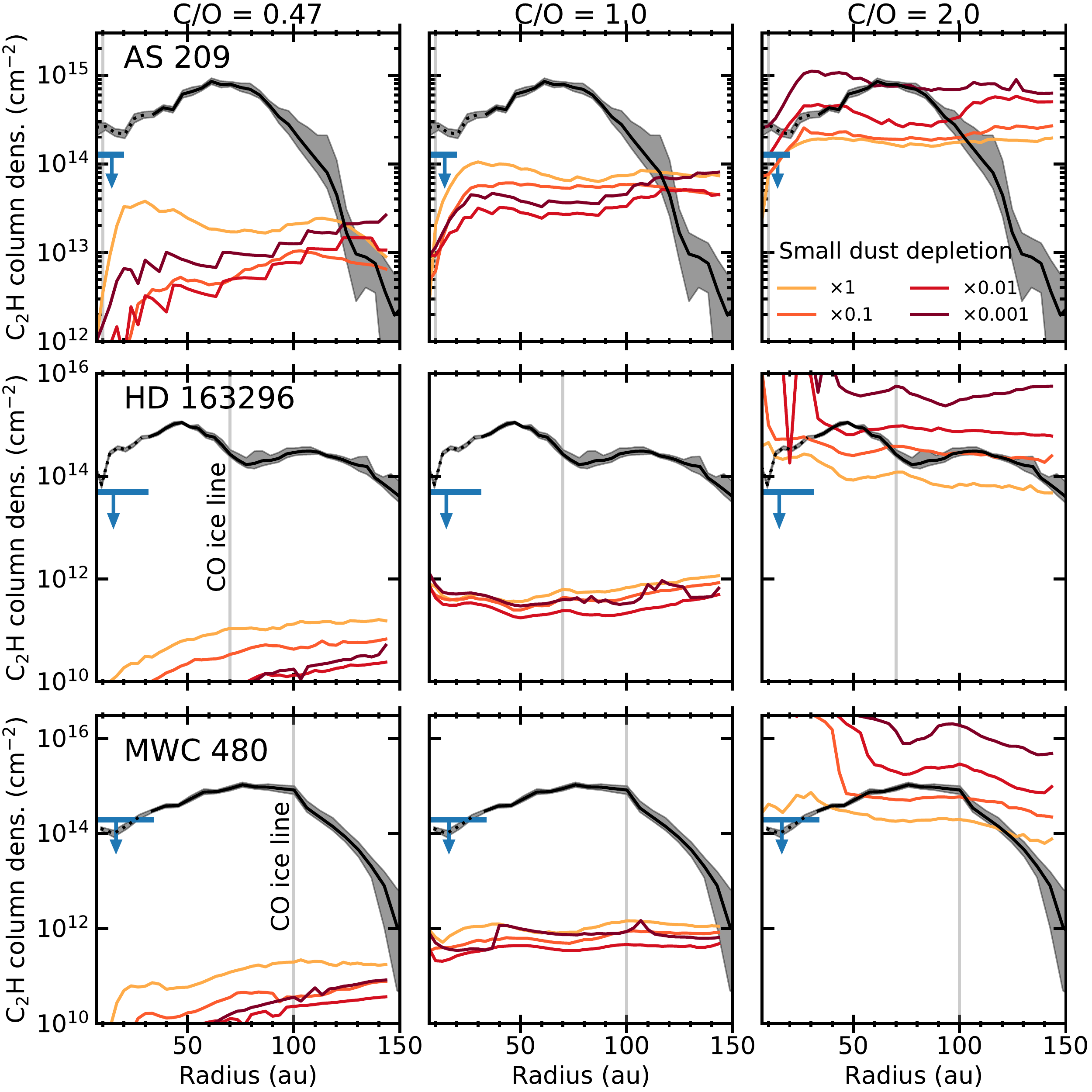}
    \caption{\ce{C2H} column densities as function of radius for AS 209 (top), HD 163296 (middle) and MWC 480 (bottom). From left to right the figures show models with C/O ratios of 0.47, 1.0 and 2.0, with the colors showing the depletion of the small dust grains, from the fiducial value (yellow) to 1000$\times$ depleted (dark red). The black line with the grey error region shows the column densities extracted from the data \citep{guzman20}, the black line transitions into a dotted line in the harder to probe inner 30 au. The blue bars and arrows show the inner disk \ce{C2H} column density upper limits derived from the line profiles (see Appendix~\ref{app:C2H_col_deriv}). The vertical grey line shows the location of the CO ice line in the models. The model \ce{C2H} column densities are integrated over both sides of the disk. High C/O ratios are necessary to reproduce the observed \ce{C2H} column densities. }  
    \label{fig:C2Hcolumns}
\end{figure*}

Figure~\ref{fig:C2Hcolumns} shows the model and observed \ce{C2H} column densities for the AS 209, HD 163296, and MWC 480 disks for C/O ratios of 0.47, 1.0, and 2.0, and the different depletion factors for the small dust grains. Two dimensional \ce{C2H} abundances structures can be found in Appendix~\ref{app:C2H_abu}, together with a discussion on the chemical behavior of the \ce{C2H}. As expected, the C/O ratio has a large effect on the \ce{C2H} column density, especially the jump from C/O = 1.0 to C/O = 2.0.

Figure~\ref{fig:C2Hcolumns} clearly shows that C/O ratios $\geq$ 2.0 are necessary to match the high, $>10^{13}$ cm$^{-2}$, \ce{C2H} column densities observed in the inner 100--150 au of the AS 209, MWC 480 and HD 163296 disks.
In all sources, the peak in the \ce{C2H} column density distribution require the signficant small-grain depletion, with the AS 209 model requiring a depletion in small-grains of 1000 over our baseline model to reproduce the \ce{C2H} column density. Maxima in the derived \ce{C2H} column have some co-locality with the first deep gap in the millimeter disk and region of low CO abundance \citep[Fig.~\ref{fig:CO_C2H_comp}][]{law20_rad, zhang20, guzman20}. However,  peaks in \ce{C2H} emission (and column) are much broader than the gaps seen in the millimeter disk. Interestingly the secondary peak around $\sim$110 au in the HD 163296 \ce{C2H} abundance structure does not line up with the second strong dust gap around $\sim85$ au. This is further explored in Sec.~\ref{ssc:Struct_C2H}. 

The \ce{C2H} column density drops sharply towards large radii ($>100$ au, $>150$ au and $>125$ au for AS 209, HD 163296 and MWC 480, respectively). A constant C/O ratio model does not share this abundance drop at this radius, implying that the C/O ratio drops around these radii in these disks.

All three disks show a decrease in the \ce{C2H} column density inwards of 50 au. The AS 209 disk models show a dip in \ce{C2H} column in this region, driven by the decreasing CO, and thus available carbon abundance in the models. In HD 163296 and MWC 480, the \ce{C2H} column density decreases in the region where the \ce{CO} abundance and thus available carbon content increases. This implies a drop in C/O ratio in the inner regions of these disks. While this drop is expected to happen at the CO ice line, it appears to happen at a smaller radius for both the HD 163296 and MWC 480 disks. This is discussed in more detail in Sec.~\ref{ssc:CO_C_O}.

In summary, a high C/O ratio, together with a strong depletion of small dust is necessary to match maxima in the derived \ce{C2H} column. A model with a constant C/O ratio and small-grain depletion is not able to match the full behavior of the \ce{C2H} column density radial profiles. A radial variation in C/O ratio and small dust content of the disk atmosphere are necessary to reproduce this behavior. Implications of this inference are discussed in Sec.~\ref{ssc:Struct_C2H}~and~\ref{ssc:CO_C_O} and for the AS 209 disk, this is studied in more detail in \citet{alarcon20}.

\section{Discussion}

\subsection{C/O ratios from \ce{C2H} in the literature}

The inferred C/O ratio associated with C$_2$H emission locations in these disks are 2.0 or higher above most of the milli-meter disk. This is higher than previously derived for other disks. \citet{Miotello2019} found that a global C/O ratio of 1.5 reproduced the flux for the sample of Lupus disks studied, significantly lower than the C/O ratios found for these disks. This discrepancy is caused by the fact that the MAPS data resolve the disk emission and reveal existing emission structure. 
In the gas with strong C$_2$H emission, a C/O ratio $\geq 2.0$ is necessary.  However,  the \ce{C2H} column density is not elevated ($>10^{14}$ cm$^{-2}$) over the entire disk.  Thus in models using disk-integrated flux  a lower, global C/O ratio, will be estimated.  This would be especially prominent for the three disks studied here as the \ce{C2H} column densities drop below $10^{12}$ cm$^{-2}$ around 150 au \citep[see,][]{guzman20}, which imply C/O ratios $< 1$ for the majority of the disk area.

In the IM Lup disk, the C/O ratio was determined by \citet{Cleeves2018} to be around 0.8, even lower than the C/O ratios found for most of the disk in \citet{Miotello2019}. When comparing the \ce{C2H} radial profiles in \citet{law20_rad} and \ce{C2H} column densities in \citet{guzman20} between the IM Lup disk and the AS 209, MWC 480 and HD 163296 disks, it is clear that the IM Lup disk has less \ce{C2H} than the other disks by more than an order of magnitude. The observationally derived \ce{C2H} column density for the IM Lup disk is $\sim 10^{13}$ cm$^{-2}$ between 70 and 230 au, which is of the same order as the AS 209 disk model column densities for a C/O of 0.47 and 1.0. The elemental composition of the IM Lup disk is thus clearly different from the three disks studied here.

The difference between the composition of the IM Lup disk and the other disks studied here could be due to the young age of the IM Lup disk \citep[][]{Mawet2012}. The process responsible for turning the volatile rich and oxygen dominated ISM composition gas into the volatile poor and carbon rich gas seen in many other disks could then still be underway \citep{Zhang2020b, Bergner2020}. This composition transition can either be the result more efficient trapping of oxygen carrying species, e.g. water, CO, and CO$_2$, as ices coating mm-sized dust particles that are confined to the mid-plane \citep{Krijt2020}, or the release of carbon from a refractory source in a uniformly depleted surface layer \citep{Bosman2021}. Alternatively, the IM Lup disk has less pronounced substructures in the millimeter disk. This could simply be a result from the young age, but it can also imply that these substructures are, at least partly, responsible for the high C/O ratios \citep{Bergin2016, Bosman2021}. 

\subsection{Structure in the \ce{C2H} column density}
\label{ssc:Struct_C2H}
While it is clear that to match the high ($> 10^{13}$\,cm$^{-2}$) column densities of \ce{C2H} in the AS 209, MWC 480, and HD 163296 disks a C/O $\geq 2$ is necessary, it is not obvious what causes the radial structures in the \ce{C2H} column density. These can be caused by either a variation of C/O ratio when the C/O ratio is below 2.0, a variation in small dust content when the C/O ratio is larger than unity or a combination of both. The effect on the \ce{C2H} column density due to changes in the C/O ratio are generally larger than those due to changes in the small dust content (Fig.~\ref{fig:C2Hcolumns}).

The regions in which a depletion of small-grains must be invoked to match the \ce{C2H} column density is wider than the millimeter dust gaps seen in these three disks. As the chemical timescales are short, $<$ 1000 yr \citep{Bosman2021}, diffusive spreading of \ce{C2H} from the millimeter gap, which happens on time-scales $> 10^4$ yr, can be ruled out. This is thus suggestive of a wider small-grain gap, around the millimeter dust gap, possibly due to the millimeter dust rings acting as a small-grain sink in a manner similar to that predicted from the midplane \citep[][]{Krijt2016}.

In the HD 163296 disk, the \ce{C2H} column density outside 70 au is consistent with an elevated C/O ratio of 2.0 and factor $\sim 10$ of small dust depletion.  At 50 au, the radius of a prominent gap in the millimeter continuum, the \ce{C2H} column density is more comparable to the model with a factor 100 depletion is small-grains. This indicates that the UV penetrates deeper into the disk around the millimeter dust gap. This is not seen in the second dust gap at 90 au, where there is a local minimum in the \ce{C2H} abundance. A possible explanation for the difference in \ce{C2H} column density can be found in the scattered light image. There is a bright ring in scattered light associated with the 66 au sub-millimeter dust ring \citep[e.g.][]{Garufi2014, Monnier2017, MuroArena2018}. This ring of abundant small dust could shadow the region of the second dust gap lowering the available UV flux in the gap and thus lowering the \ce{C2H} column density.  

The \ce{C2H} column density in the MWC 480 disk peaks around 70 au, the location of the millimeter continuum dust gap \citep{Long2019, Liu2019, law20_rad}. Similar to the gap at 50 au in the HD 163296 disk, high UV penetration at the gap location -- again suggestive of small-grain depletion -- is necessary.

The need for additional small dust depletion on top of the fiducial dust models that already have a degree of dust depletion in the surface layers implies gas-to-dust ratios in the surface layers of at least 10000  and up to $10^{6}$ at the location of the \ce{C2H} emission peak in AS 209 \citep{zhang20}. It is thus possible that other processes are increasing the \ce{C2H} abundance above the levels, predicted by the chemical models. One intriguing possibility is that strong vertical or radial mixing is operating, which could cycle \ce{C2H2} and other larger carbon species from more shielded regions into the 25-75 au region. This would lead to an out of equilibrium chemistry with a higher \ce{C2H} abundance. This could be consistent with a large scale meridional flow into the gaps and would be an interesting subject of further study.

\subsection{A link between CO abundance and C/O}
\label{ssc:CO_C_O}
It has been suggested that a low CO abundance and a high C/O ratio have a common dynamical or chemical origin, which should cause a correlation between CO abundance and C/O ratio \citep{Bergin2016}. However, a relation between \ce{C2H} flux and CO abundances, as traced by \ce{^{13}CO} over millimeter dust flux has not been found \citep{Miotello2019}.
The relation has not been studied as a function of radius in a disk, where, for example, the release of CO at the CO ice line is expected to have a strong impact on the gas-phase C/O ratio \citep{Oberg2011}.


For the MWC 480 and HD 163296 disks, the regions within and outside of the \ce{CO} ice line are resolved by the observations. The emission maps of C$^{18}$O (and the derived CO column) do not exhibit a strong response at the CO ice line location predicted by the detailed disk physical models. Most directly, the CO abundance at the CO ice line does not return to the ISM level \citep[e.g. 10$^{-4}$ relative to H$_2$;][Fig.~\ref{fig:CO_C2H_comp}]{zhang20}. This is consistent with a lack of the expected \ce{C2H} column density drop at the \ce{CO} ice line. Without the return of CO, the C/O ratio does not drop towards unity, but stays high. Closer to the star, there is an increase in the CO abundance, the CO abundance reaching greater than ISM levels at 50 au in MWC 480 and at 30 au in HD 163296.

The increase in CO abundance at these radii is accompanied by a drop \ce{C2H} column density see Fig.~\ref{fig:CO_C2H_comp}, consistent with a drop in C/O ratio (Fig.~\ref{fig:C2Hcolumns}) due to the return of the oxygen carried by CO into the gas. As mentioned the location that the CO abundance rises is not where the CO ice line is expected, based on the thermal structure in the models \citep{zhang20, Calahan20b}. Temperatures in the region where the CO abundance rises are between 25 and 30 K in the standard MAPS models \citep{zhang20}. These temperatures are in line with modeling by \citet{Calahan20b} for HD 163296 and empirical temperature measurements from the CO isotopologue emission \citep{law20_surf}. Furthermore the location where the \ce{C2H} column drops below the C/O=2 models is at smaller radii than the rise in CO abundance. This implies that when the CO abundance rises to close to the ISM value, then the  C/O ratio drops.

If the rise of CO abundance is due to CO coming off the grains, this would imply a higher binding energy than the 850 K \citep[$\sim$ 22 K iceline temperature, e.g.][]{Harsono2015} assumed in the models, or be due to a different physical effect such as large grains drifting, and only fully releasing their icy CO mantle further in. Exact effects are going to be very disk dependent. A higher CO desorption temperature was also inferred for CO in TW Hya \citep{Zhang2017}. This could be indicative for CO as desorbed on \ce{H2O} or \ce{CO2} mixture \citep{Noble2012, He2014}. In the case of HD 163296 and MWC 480, in which the CO content of the grains is expected to be high, this would indicate that the vast majority of the $\sim$ 100 monolayers of CO are all in high binding energy sites. 
To move the ice line by $\sim$40 au due to fast radial drift grains between 1 and 10 cm are necessary to reach the right drift speeds and desorption speed \citep{Piso2015}. This is larger than the 3 mm grains that \citet{sierra20} need at these radii to fit multiband continuum data. 

Similar relation between CO and C2H abundance is seen in the outer disk of AS 209, where outside 100 au the CO abundance increases, and the \ce{C2H} column density drops, again implying a low C/O ratio when the CO abundance is high. HD 163296 and MWC 480 also show a drop in \ce{C2H} column density and inferred C/O ratio outside $\sim$120 and $\sim$150 au respectively. This is not paired with an increase in CO abundance, however. This implies that a high C/O ratio happens only in the regions with low CO abundance. However, something in addition to low CO abundance is needed to explain a high C/O ratio. This could be related to the presence of an additional carbon source, such as the photo-ablation of carbonaceous grains \citep{Bosman2021}.

\subsection{Composition of planet building material}

To properly predict the composition of a planet forming at a certain location, it is necessary to know the elemental composition of the building materials. For any planet forming outside of 30 au in the AS 209, HD 163296, and MWC 480 disks, such as those inferred in the various gaps in these systems, it is now possible to make some inferences on the composition of the gas and solids that will go into these putative planets. For this discussion we will assume that the bulk of the gas mass is accreted late in the disk lifetime such that the composition of the disk gas inferred here is representative of the gas accreted onto the planet. If planets accrete their gas mass early in the disk lifetime, when accretion onto the disk still delivers significant pristine material, planetary composition would be very similar to that of the host star.

The \ce{C2H} emission in these three disks is strong, but an emission surface height cannot be derived unambiguously over the entire disk \citep{law20_surf}. This either implies that the line is optically thin, and thus too weak to measure a height or that it comes from below the \ce{^{13}CO} surface, which results in a measured height that is consistent with the disk midplane. The hyperfine ratios, from which the \ce{C2H} column densities have been derived, show that the lines are optically thick. This means that the \ce{C2H} lines come from deeper than the \ce{^{13}CO} emission. This is consistent with the disk regions where the \ce{C2H} height can be derived \citep{law20_surf} and is in agreement with the models, that have most of the column density between 1 and 2 scale-heights \citep[see Appendix~\ref{app:C2H_abu}, the scale-heights in these disks at 10-200 au vary between z/r of 0.06 and 0.1 depending on the disk, ][]{zhang20}. At this height, mixing with the midplane is efficient and happens on 10~kyr time-scales for turbulent viscosity coefficients of $\alpha \approx 10^{-3}$ \citep{Ciesla2010}. As such, it is unlikely that strong vertical elemental abundance gradients exist, so the high C/O abundances and low O/H abundances necessary to explain the \ce{C2H} emission likely hold for the disk midplane as well. Furthermore, if accretion onto a planet happens through meridional flows, the gas accretion onto the planet will actually sample the vertical height from where the \ce{C2H} emission originates \citep{Tanigawa2012, Morbidelli2014}; such flows have been observed in the higher layers of the disk with \ce{^{12}CO} in the HD 163296 disk by \citet{Teague2019}. As these are the flows that feed the planet, gas from the \ce{C2H} layer emitting layer is expected to contribute significantly to the planetary accretion after gap opening \citep{Morbidelli2014,Cridland2020}.


The amount of small-grains in the disk atmosphere, that can accrete with the gas onto the planet, is strongly reduced from the ISM level of 1\% of the total mass. This happens both globally in the disk surface due to grain growth and settling, and locally, as the inferred \ce{C2H} column densities necessitate more UV penetration in certain disk locations. The combined low CO abundance, and depleted small dust thus imply that most of the ISM carbon and oxygen, both volatile and refractory, is not present in gas accreting onto planets that have reached pebble isolation mass.

Gas giants that accrete most of their atmospheres after reaching the pebble isolation mass thus obtain envelopes, with substellar C/H and O/H, and high ($\gg1.0$) C/O ratio compositions, unless efficient release of volatiles from the core or late enrichment by planetesimals takes place. These potential planets are formed at large radii, where CO can be depleted efficiently ($>$ 50 au radii for HD 163296 and MWC 480). If these planets do not migrate from their formation location, they are prime targets for composition analysis by direct imaging with the James Webb Space Telescope, the most notable example being the planets in the HR 8799 system. Substellar carbon and oxygen abundances and a high C/O composition would distinguish planets formed through pebble or core accretion from planets formed as a result of a gravitational instability. The latter are expected to have a stellar or superstellar C/H and O/H ratios with a stellar or substellar C/O ratio \citep[e.g.][]{Ilee2017}.

Compositions of planets at large radii so far do not show the signature predicted from the disk observations. $\beta$ Pic b has a low derived C/O ratio, which is expected to be close or below the stellar C/O ratio \citep{Gravity2020}. This rules out a formation scenario in which one of the $> 40$ au putative planets in the disks studied here migrates inward to 9 au \citep{Lagrange2020}, unless this planet accreted $>$40 M$_\oplus$ of oxygen-rich ice, or mixed a similar amount of ice from the core into the atmosphere \citep{Gravity2020}. Three other large radii ($>50$ au) planets have their water abundance measured \citep[$\kappa$ And b, HR8799 b and HR8799 c][]{Todorov2016, Lavie2017, Madhusudhan2019}. The water abundances are higher than the disk atmosphere gas, ice and refractory oxygen abundances combined, by a factor 2 to 100, for the three disks studied here. If these planets are the successors of the putative planets in these disks, a significant amount of disk water ice needs to make it into the atmosphere, at least 30 $M_\oplus$ for $\kappa$ And b and HR 8799 b, and around 100 $M_\oplus$ for HR 8799 c \citep{Madhusudhan2019}. This would imply very massive cores (60 - 200 $M_\oplus$) and very efficient core mixing, 6-20 $M_\oplus$ Myr$^{-1}$ planetesimal accretion rates for a 10 Myr disk age, or a gravitational instability origin from moderately (factor 1-3) water ice enhanced core.  The low water abundance measured for many transiting exoplanets is comparable to the low total oxygen abundance in the gas at large ($>50$ au) radii in the AS 209, HD 163296 and MWC 480 disks \citep[][]{Madhusudhan2019, Welbanks2019}. This could point at an outer disk origin for these hot Jupiters, however without constraints on the C/O ratio this remains speculation. A high C/O $\gtrsim 1.0$ would strengthen the case for an outer disk origin for hot Jupiters.

\section{Conclusions}
We have studied the behavior of the \ce{C2H} column densities above the millimeter pebble disks orbiting AS 209, MWC 480, and HD 163296, and compared these to \ce{C2H} column densities to source specific thermochemical models calculated using DALI \citep{Bruderer2012,Bruderer2013} with an extended gas-grain chemical network. We find that high ($\sim 2.0$) C/O ratios are necessary to explain the strong \ce{C2H} emission. The high column densities ($>10^{15}$ cm$^{-2}$) imply that there is an additional effect at play. One option is more efficient UV penetration, leading to even higher \ce{C2H} column densities. The high, but localized, \ce{C2H} column densities explain why these resolved observations need locally higher C/O ratios than global models, like those from \citet{Miotello2019}.

Furthermore, a relation between the C/O ratio and CO abundance appears to exist, with a higher CO abundance leading to a lower C/O ratio.  This is especially the case when looking at the inner 100 au of the HD 163296 and MWC 480 disks. The transition to an enhanced CO abundance and diminished C/O ratios seems not to happen at the CO ice line, but further in. 

The C/O ratio and the gas-to-dust ratio are elevated above stellar and ISM-levels (respectively) along the radial extent of the pebbled disk, except in the regions where the CO abundance is high (inner $\sim$30 au for HD 163296 and MWC 480). This includes the regions that most of the dust gaps are located. The gas accreting onto the possibly forming planets is thus strongly depleted of volatile and refractory carbon and oxygen, with a strongly super-stellar C/O ratio. In the absence of core-envelope mixing and envelope enrichment by planetesimal collisions, this would lead to a high C/O but generally volatile depleted giant planets. Comparing this prediction to the composition of gas giants at large orbits, such as will be inferred by the \textit{James Webb Space Telescope}, will be able to point at the origin of these large orbit planets.

\acknowledgments

This paper makes use of the following ALMA data: ADS/JAO.ALMA\#2018.1.01055.L. ALMA is a partnership of ESO (representing its member states), NSF (USA) and NINS (Japan), together with NRC (Canada), MOST and ASIAA (Taiwan), and KASI (Republic of Korea), in cooperation with the Republic of Chile. The Joint ALMA Observatory is operated by ESO, AUI/NRAO and NAOJ. The National Radio Astronomy Observatory is a facility of the National Science Foundation operated under cooperative agreement by Associated Universities, Inc.

A.D.B., E.A.B., F.A., and K.I.Ö. acknowledge support from NSF AAG grant No. 1907653.
K.Z. acknowledges the support of the Office of the Vice Chancellor for Research and Graduate Education at the University of Wisconsin – Madison with funding from the Wisconsin Alumni Research Foundation. 
K.Z., J.B.B., I.C., J.H. \& K.R.S. acknowledge support from NASA through the NASA Hubble Fellowship grants, HST-HF2-51401.001, \#HST-HF2-51429.001-A, HST-HF2-51405.001-A, \#HST-HF2-51460.001-A \& \#HST-HF2-51419.001 awarded by the Space Telescope Science Institute, which is operated by the Association of Universities for Research in Astronomy, Inc., for NASA, under contract NAS5-26555.
M.L.R.H. acknowledges support from the Michigan Society of Fellows
K.I.\"O. acknowledges support from the Simons Foundation (SCOL No. 321183). 
V.V.G. acknowledges support from FONDECYT Iniciaci\'on 11180904 and ANID project Basal AFB-170002.

C.W. acknowledges financial support from the University of Leeds, STFC and UKRI (grant numbers ST/R000549/1, ST/T000287/1, MR/T040726/1).
Y.A. acknowledges support by NAOJ ALMA Scientific Research Grant Code 2019-13B, and Grant-in-Aid for Scientific Research 18H05222 and 20H05847.
S. M. A. and J. H. acknowledge funding support from the National Aeronautics and Space Administration under Grant No. 17-XRP17 2-0012 issued through the Exoplanets Research Program.
A.S.B acknowledges the studentship funded by the Science and Technology Facilities Council of the United Kingdom (STFC).
G.C. is supported by NAOJ ALMA Scientific Research grant Code 2019-13B.
L.I.C. gratefully acknowledges support from the David and Lucille Packard Foundation and Johnson \& Johnson's WiSTEM2D Program.
J.D.I. acknowledges support from the Science and Technology Facilities Council of the United Kingdom (STFC) under ST/T000287/1.
C.J.L. acknowledges funding from the National Science Foundation Graduate Research Fellowship under Grant No. DGE1745303.
R.L.G. acknowledges support from a CNES fellowship grant.
Y.L. acknowledges the financial support by the Natural Science Foundation of China (Grant No. 11973090).
F.L. \& R.T. acknowledge support from the Smithsonian Institution as a Submillimeter Array (SMA) Fellow.
F. M. acknowledges support from ANR of France under contract ANR-16-CE31-0013 (Planet-Forming-Disks)  and ANR-15-IDEX-02 (through CDP "Origins of Life").
H.N. acknowledges support by NAOJ ALMA Scientific Research Grant Code 2018-10B and Grant-in-Aid for Scientific Research 18H05441.
T.T. is supported by JSPS KAKENHI Grant Numbers JP17K14244 and JP20K04017.
Y.Y. is supported by IGPEES, WINGS Program, the University of Tokyo.
\software{SciPy \citep{Virtanen2020},  NumPy \citep{van2011numpy}, Matplotlib \citep{Hunter2007}.}

\bibliographystyle{aasjournal}
\bibliography{Lit_list, MAPS_bib}

\appendix

\section{\ce{C2H} column in the inner 30 au}
\label{app:C2H_col_deriv}
\begin{figure}
    \centering
    \includegraphics[width = \hsize]{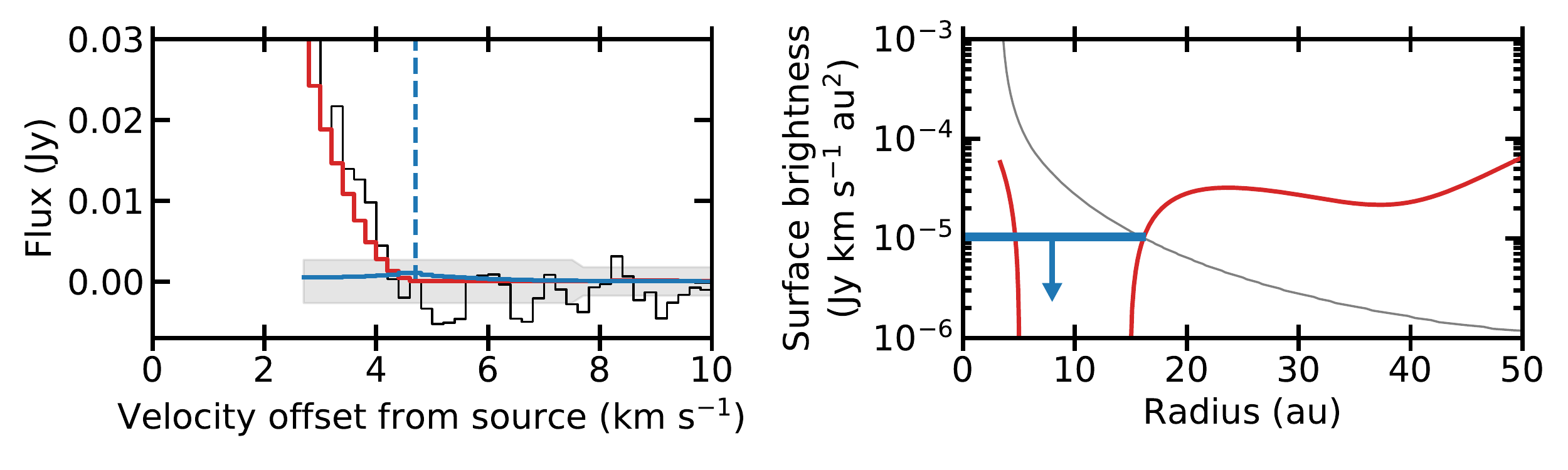}
    \caption{Left:} Redshifted line wing of the \ce{C2H} $N$=3--2; $J=$7/2-5/2, $F=$4--3 transition in AS 209 (black). The red line shows a model fit to the flux at $>$2.6 km s$^{-1}$ from source velocity, corresponding to $<$55 au for AS 209 and the blue line shows the profile of the flux upper limit in the inner region. Right: The fitted radial profile (red) and flux upper limit (blue) derived from the data. The grey line shows relation between the extend of an emitting area and the minimal, constant flux that would be detectable from that area given the noise in the data. This is our sensitivity limit estimate. The point where the fitted radial profile dips under this sensitivity limits is where we set the flux upper limit for the inner most region. 
    \label{fig:linewing}
\end{figure}

To set limits on the \ce{C2H} column from the inner regions of the disk, we use the kinematic information in the \ce{C2H} line profile to determine the flux that is actually coming from the inner disk. First we filter out flux that seems to be emitted from small radii in the image, but is  instead flux from larger radii, smeared to the inner regions by the beam. To accomplish this we use the method outlined in \citet{bosman20_inner20au} to extract spectra from the image cube and then transform these inner disk spectra into radial profiles using the kinematical information assuming the disk is in Keplerian rotation. Instead of using both the red and blue-shifted line wings as done for CO, we only use the uncontaminated line wing for \ce{C2H}. Radial profiles are created for the four strong components of the \ce{C2H} $N$=3--2 transition ($J$=7/2--5/2, $F$=4--3; $J$=7/2--5/2, $F$=3--2; $J$=5/2--3/2, $F$=3--2 and $J$=5/2--3/2, $F$=2--1). The weaker $N$=3--2, $J$=5/2--3/2, $F$=2--2 line does not contaminate the spectrum of the nearest strong line. Figure~\ref{fig:linewing} shows the redshifted wing of the $N$=3--2; $J=$7/2-5/2, $F=$4--3 transition together with the extracted radial profile and inner region upper limit for AS 209. 

\begin{figure}
    \centering
    \includegraphics[width = \hsize]{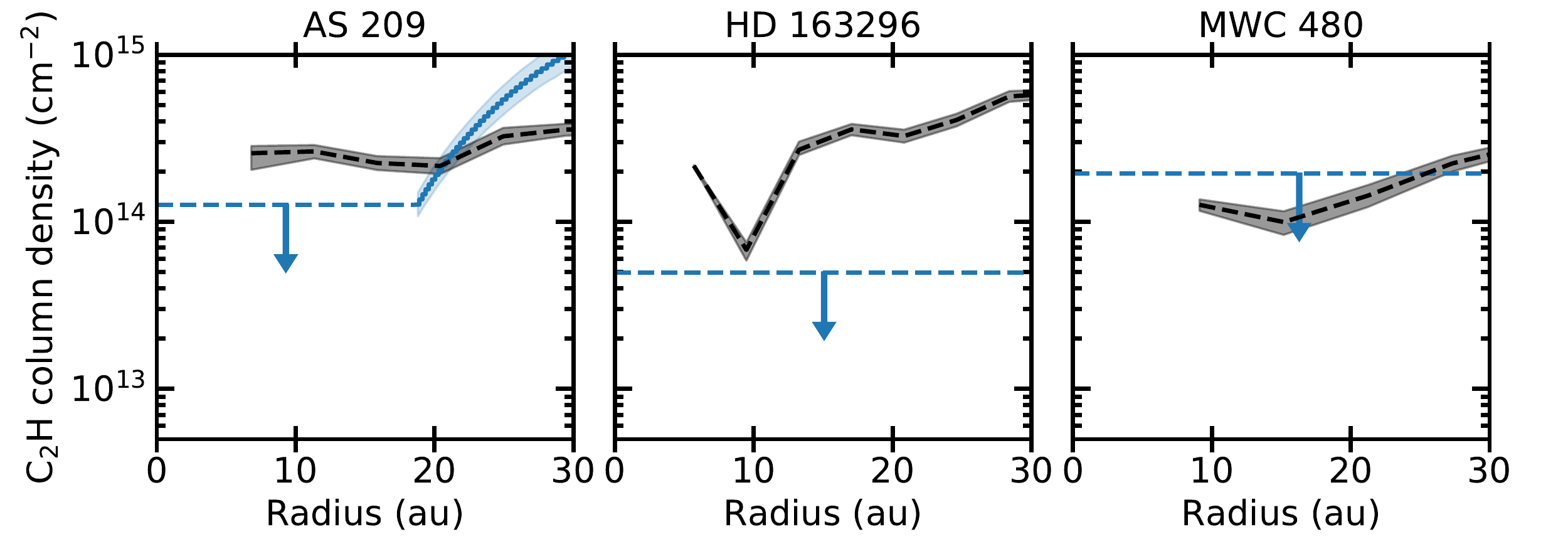}
    \caption{Comparison of the image derived \ce{C2H} column \citep[black,][]{guzman20} and the column density derived from the line profiles (blue) in the inner 30 au for the AS 209, HD 163296 and MWC 480 disk. The analysis in the image plane is overestimating the \ce{C2H} column in the inner disk.  }
    \label{fig:C2H_inner_comp}
\end{figure}

To account for the loss of sensitivity by only having one line-wing per fit, we increase the radial grid spacing in the radial profile fitting to 10 au. The emission is weak and within 30 au we are  mostly constrained to estimating upper limits. Only for the AS 209 disk can some flux be ascribed to the 20-30 au region. The radial extend as well as the flux upper limit derived are taken from the radial profiles and these are converted into a column upper limit for the inner disk. To convert flux into column densities we follow \citet{bergner20} taking the quantum values for \ce{C2H} from the JPL database (see \citet{guzman20}). To convert flux to column, gas temperature and continuum emission temperature are necessary. For the gas temperature we assume the temperature from \citet{law20_surf} for \ce{^{12}CO} $J$= 2--1 at 30 au and for the continuum we take the brightness temperature at 30 au from \citet{sierra20}, using disk center values changes derived columns less than a factor 2. Values are given in Table.~\ref{tab:temp}. We also assume LTE and that the line width is fully thermal. Fig.~\ref{fig:C2H_inner_comp} compares the derived limits from the line profiles to the columns derived from the images.

\begin{table}[]
    \caption{Temperatures used for the \ce{C2H} column derivation}
    \centering
    \begin{tabular}{l r r}
    \hline
    \hline
    Source  & Gas temperature (K) & Dust emission temperature (K) \\
    \hline
    AS 209      &   30.2   &  12.4 \\
    HD 163296   &   53.8   &  22.7 \\
    MWC 480     &   69.1   &  25.5 \\
    \hline
    \end{tabular}
    \label{tab:temp}
\end{table}

\section{Tracing the gaseous volatile content with CO}
\label{app:CO_cols}
\begin{figure*}
    \centering
    \includegraphics[width = \hsize]{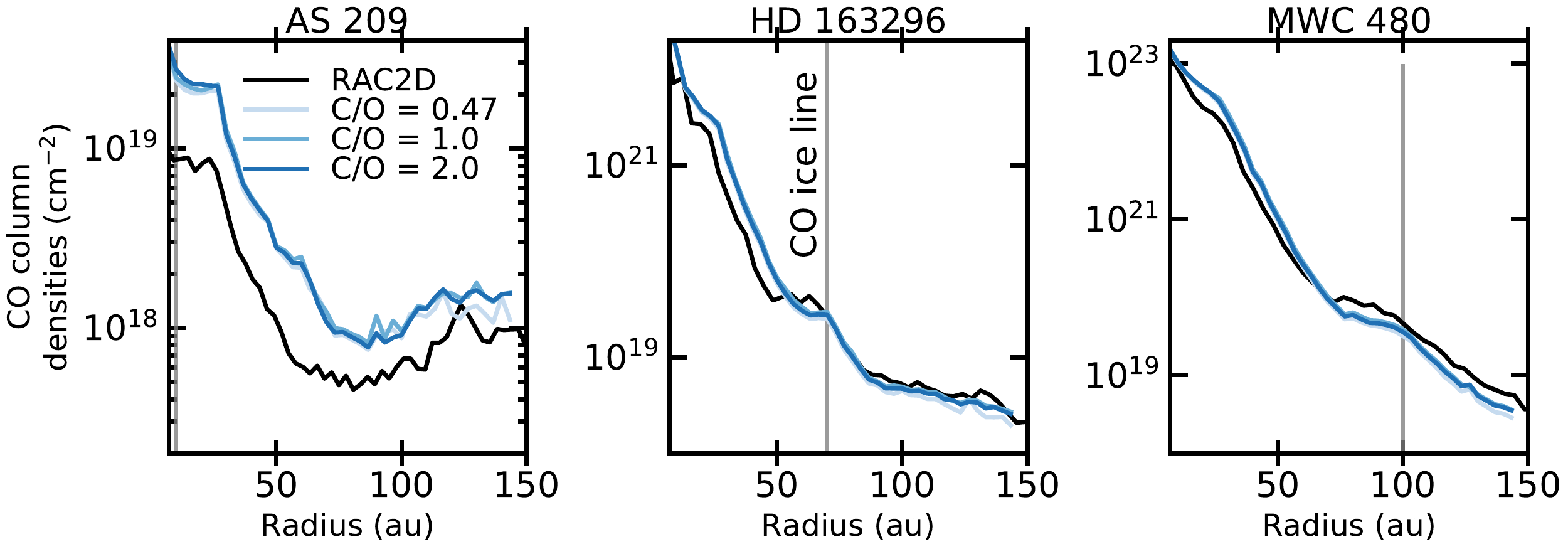}
    \caption{ CO column densities for the fiducial dust models for different C/O ratios (blue) compared to the derived CO column densities from the data \citep[black, ][]{zhang20}. Vertical grey lines show the CO ice line location extracted from the RAC2D models \citep{zhang20, Calahan20b}. }
    \label{fig:CO_columns}
\end{figure*}
Figure~\ref{fig:CO_columns} compares the CO column densities derived by \citet{zhang20} and the CO column densities extracted from the gas-grain network for the three different C/O ratios tested here for the thermochemical models with undepleted small dust. The CO column densities in the DALI models match well for models of the MWC 480 and HD 163296 disks, but the AS 209 DALI models produce higher CO column densities than in the AS 209 RAC2D model.

The AS 209 RAC2D model has a significant amount of carbon in \ce{CO2} ice, while in the DALI models \ce{CO2} is barely formed, leaving more carbon in gas-phase CO near the midplane. This is due to the more simplified implementation of the grain surface chemistry in RAC2D \citep{Du2014}. This difference is only seen in the AS 209 model due to the larger X-ray luminosity and lower UV luminosity in the AS 209 model compared to the HD 163296 and MWC 480 models. In the latter two models, the low X-ray flux and higher UV flux suppresses the formation of \ce{CO2} in the RAC2D models, putting them in line with the DALI models. 

Lowering the carbon and oxygen abundances in the DALI AS 209 models by a factor of 2-3 within 120 au would lead to a better match of the DALI CO column to the CO column that is used to reproduce the observed emission in the RAC2D models. This drop in carbon and oxygen abundance would only very slightly lower the \ce{C2H} abundance and would not impact our conclusions on the C/O ratios of these disks. Therefore we will continue to use the CO depletion factors as derived for AS 209 in \citet{zhang20}.

The small variation between the DALI models with different C/O ratios showcases the chemical stability of CO. In both oxygen and carbon-rich environments CO incorporates either the vast majority of the carbon, or of the oxygen, whatever is in short supply.  CO remains the most abundant observable molecular carrier of elemental C and O. The CO column is thus a good tracer for the gaseous volatile content at a large range of C/O ratios.

\section{\ce{C2H} abundance structures}
\label{app:C2H_abu}
\begin{figure*}
    \centering
    \includegraphics[width = 0.7\hsize]{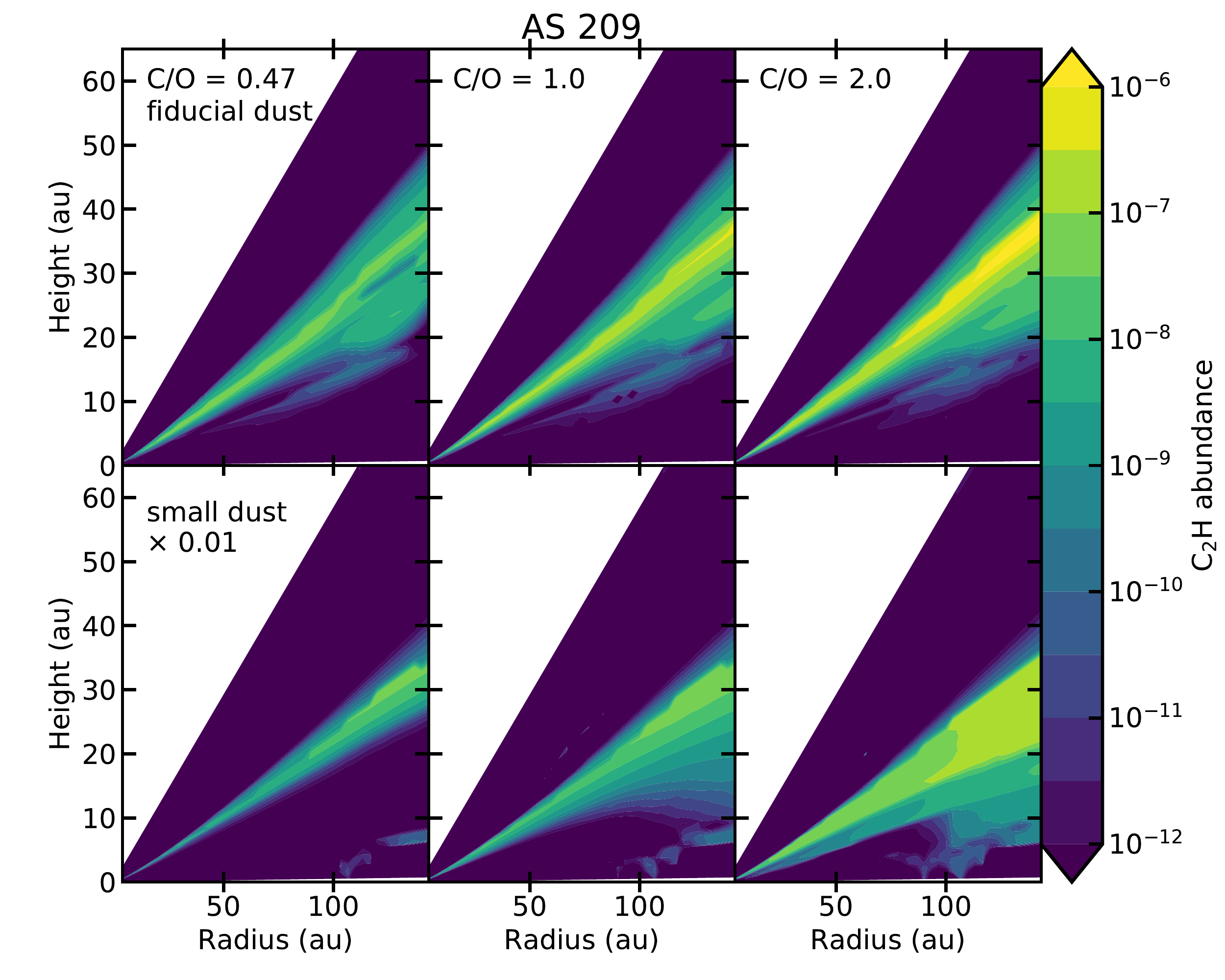}
    \caption{AS 209 \ce{C2H} abundance structure for the fiducial small dust (top) and factor 100 depleted small dust (bottom) for C/O ratios of 0.47, 1.0 and 2.0 (left to right). }
    \label{fig:as209_C2H}
\end{figure*}

\begin{figure*}
    \centering
    \includegraphics[width = 0.7\hsize]{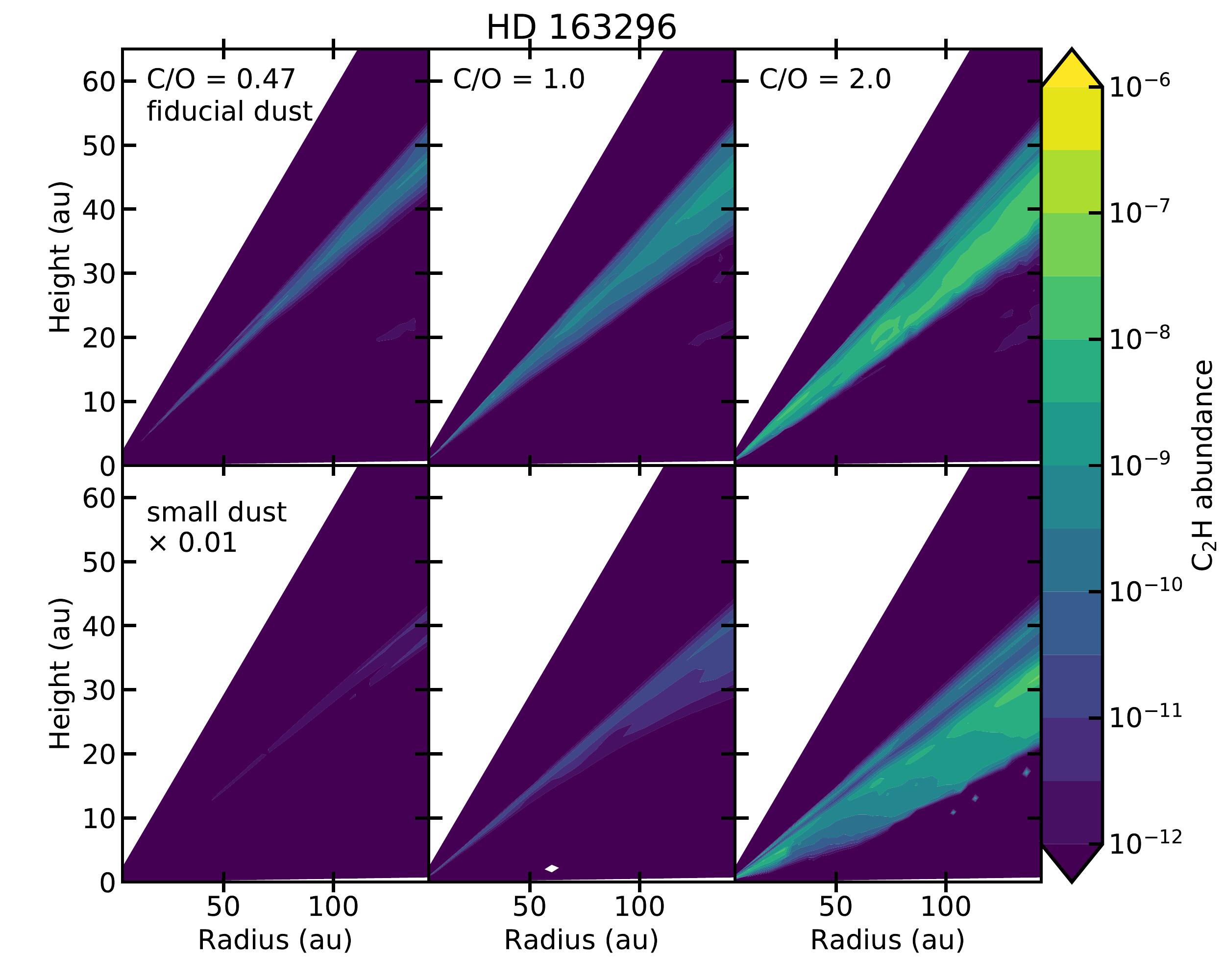}
    \caption{As Fig.~\ref{fig:as209_C2H}, but for HD 163296. }
    \label{fig:hd163296_C2H}
\end{figure*}

\begin{figure*}
    \centering
    \includegraphics[width = 0.7\hsize]{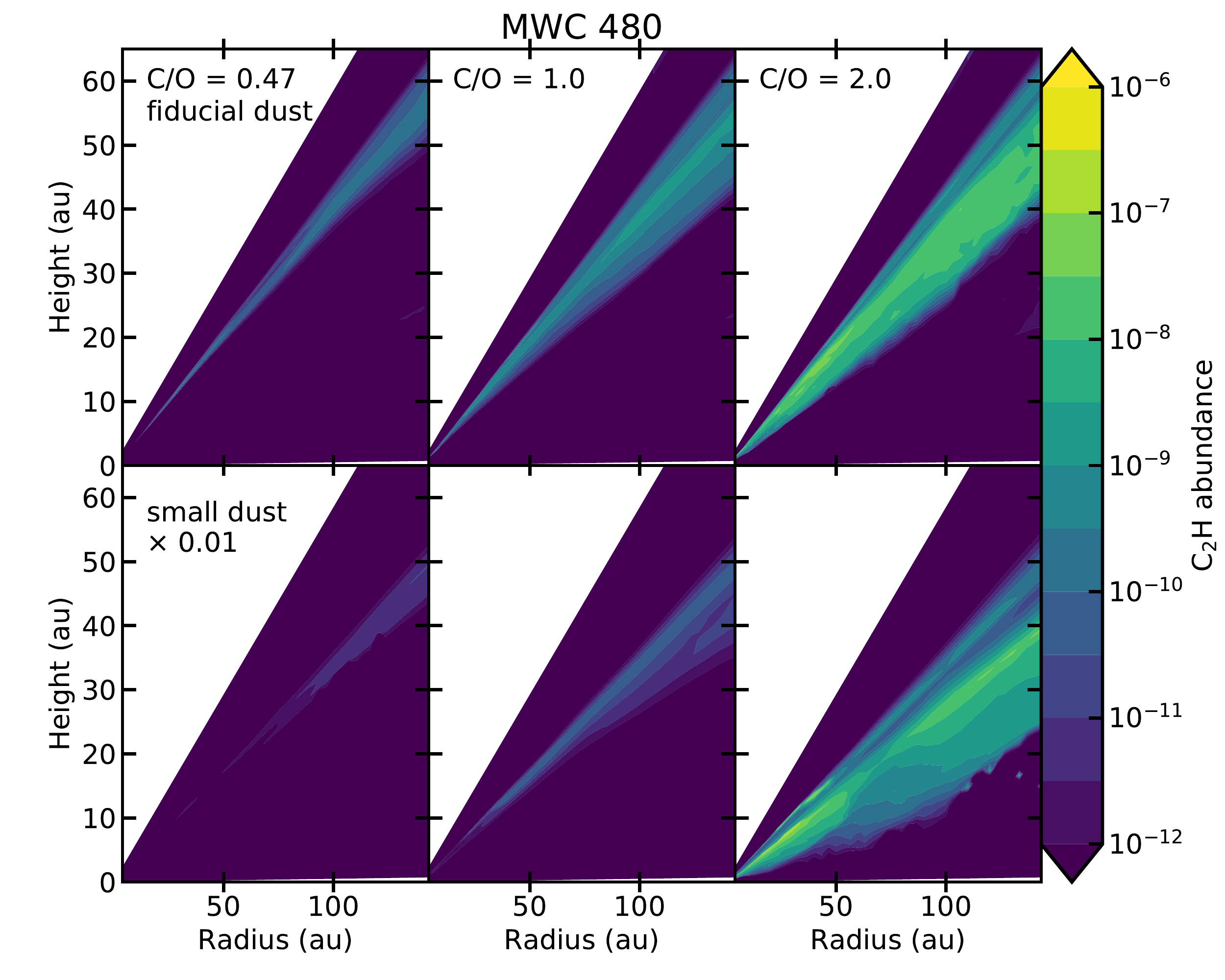}
    \caption{As Fig.~\ref{fig:as209_C2H}, but for MWC 480. }
    \label{fig:mwc480_C2H}
\end{figure*}

Figures~\ref{fig:as209_C2H},~\ref{fig:hd163296_C2H}~and~\ref{fig:mwc480_C2H} show the \ce{C2H} abundance structures for some of the AS 209, HD 163296 and MWC 480 models that feed into the column densities shown in Fig.~\ref{fig:C2Hcolumns}. There is a clear difference between the AS 209 disk and the MWC 480 and HD 163296 disks at low C/O ratios, with a higher \ce{C2H} abundance in the AS 209 disk. This is driven by different levels of UV and X-ray flux in these sources. The high X-ray flux of AS 209 allows for more efficient destruction of CO, followed by \ce{CH4} and \ce{H2O} formation. The lower gas-to-small-dust ratio in the surface layer produces less UV penetration, so \ce{CH4} and \ce{H2O} are not quickly reverted back to \ce{CO}: this allows for higher \ce{C2H} abundances at low C/O ratios. This also explains why the \ce{C2H} column density strongly drops in AS 209 for a C/O ratio of 0.47 when the small dust is depleted, the deeper penetrating UV destroys the \ce{CH4} and \ce{H2O} created by X-rays, increasing the amount of atomic oxygen in the gas and lowering the amount of carbon not in CO. The low X-ray flux and larger gas-to-dust ratios in the fiducial HD 163296 and MWC 480 disk models more efficiently force the carbon into CO. Only when carbon is more abundant than oxygen, does the \ce{C2H} production take off.

Changes in the predicted \ce{C2H} column densities due to modifications of the small dust distribution are smaller than the effect of varying the C/O ratio in all three disk models. For C/O $\leq$ 1.0, the \ce{C2H} column density is not strongly dependent on the small-grain column and no uniform change with a varying small-grain content is seen. At C/O = 2, the models with different small-grain abundances in the surface layers clearly follow the small-grain abundance.

This nonlinear interplay between the small dust abundance and C/O ratios is linked to the carbon source for the \ce{C2H}. At C/O $\leq$ 1.0, the carbon for \ce{C2H} needs to be released from CO. As CO efficiently self-shields, the amount of CO that is dissociated does not strongly depend on the continuum UV penetration, and is thus not strongly dependent on the small-grain abundance. At high C/O, the carbon comes from \ce{CH4} and other hydrocarbons deeper into the disk. Increasing UV penetration will make it easier to extract carbon from these species, and thus increases the \ce{C2H} abundance \citep[see][for a more complete discussion]{Bosman2021}.

\clearpage

\end{document}